\documentclass[aps,prd,a4paper,twocolumn,amsmath,showpacs,superscriptaddress,nofootinbib,preprintnumbers]{revtex4-1}
\usepackage{graphicx,ulem}
\usepackage{longtable}
\usepackage{float}
\usepackage{dcolumn}
\usepackage{graphics,epsfig}
\usepackage{amsmath,amssymb,latexsym,mathrsfs}
\usepackage{bm}
\usepackage{color}

\newcommand{\iso}{\mathcal{S}}
\newcommand{\reff}[1]{(\ref{#1})}
\newcommand{\Neff}{N_\mathrm{eff}}

\def\he4{$^4$He}
\def\h2{$^2$H}

\begin{document}
% Use the \preprint command to place your local institutional report
% number in the upper righthand corner of the title page in preprint mode.
% Multiple \preprint commands are allowed.
% Use the 'preprintnumbers' class option to override journal defaults
% to display numbers if necessary
%\preprint{}

%Title of paper
\title{Planck constraints on neutrino isocurvature density perturbations}

% repeat the \author .. \affiliation  etc. as needed
% \email, \thanks, \homepage, \altaffiliation all apply to the current
% author. Explanatory text should go in the []'s, actual e-mail
% address or url should go in the {}'s for \email and \homepage.
% Please use the appropriate macro foreach each type of information

\author{Eleonora Di Valentino}
%\email[]{Your e-mail address}
\affiliation{Physics Department and INFN, Universit\`a di Roma 
	``La Sapienza'', Ple.\ Aldo Moro 2, 00185, Rome, Italy}

\author{Alessandro Melchiorri}
\affiliation{Physics Department and INFN, Universit\`a di Roma 
	``La Sapienza'', Ple.\ Aldo Moro 2, 00185, Rome, Italy}
\date{\today}

\begin{abstract}
The recent Cosmic Microwave Background data from the Planck satellite experiment,
when combined with HST determinations of the Hubble constant, are compatible 
with a larger, non-standard, number of relativistic degrees of freedom at recombination,
 parametrized by the neutrino effective number $N_{eff}$. 
In the curvaton scenario, a larger value for $N_{eff}$ could arise from a non-zero
neutrino chemical potential connected to residual neutrino isocurvature density
(NID) perturbations after the
decay of the curvaton field, which component is parametrized by the
amplitude $\alpha^{NID}$. Here we present new constraints on $N_{eff}$ and $\alpha^{NID}$
from an analysis of recent cosmological data. We found that the Planck+WP dataset does not
show any indication for a NID component, severly constraining its amplitude,
 and that current indications for a non-standard $N_{eff}$ are further relaxed.
\end{abstract}

% insert suggested PACS numbers in braces on next line
\maketitle

\section{Introduction}\label{intro}

The recent measurements of the Cosmic Microwave Background anisotropies provided by
the Planck experiment have drastically improved our knowledge about the inflationary
paradigm (see e.g., \cite{Ade:2013lta}). 
In particular, several inflationary models have been ruled out and the
overall picture presented by Planck is perfectly consistent with purely adiabatic 
and gaussian primordial perturbations.

On the other hand, the recent Planck data is also showing some interesting anomaly or
tension that, albeit at small confidence level, is clearly worthwile of further investigation.

For example, the Planck data is well compatible with a larger value for
the number of relativistic degrees of freedom at recombination than what is 
commonly expected in the standard scenario (\cite{Ade:2013lta}).

Let us remind here that the energy density of relativistic particles in cosmology 
at the epoch of recombination is given by:
\begin{equation}
\rho_r= (1+N_{eff} \frac{7}{8} \Big( \frac{4}{11} \Big)^{4/3}) \rho_\gamma ,
\end{equation}
where $N_{eff}$ is the effective number of neutrinos and 
$\rho_\gamma$ is the CMB photon energy density.

It is worthwile to point out that not only additional relativistic species would
affect the value for $N_{eff}$ but also other different neutrino properties,
as a non-zero chemical potential, would change it from the standard
value of $N_{eff}=3.046$ (see \cite{Mangano:2005cc}).

In practice, the $N_{eff}$ effective parameter covers a wide range of physical
phenomena and it is therefore extremely important to check for its consistency 
with the standard expectation.

Interestingly enough, the recent Planck data does show some indication
for a non standard $N_{eff}$. For example, in the analysys of \cite{najla},
a value of $N_{eff} = 3.71 \pm 0.40$ at $68 \%$ c.l. 
from the Planck CMB data alone is reported. More importantly, when the Planck data is combined
with the measurements of the Hubble constant from \cite{hst} the 
constraint becomes $N_{eff} = 3.63 \pm 0.27$ at $68 \%$ c.l., i.e. an indication
for a non standard value at more than $95 \%$ c.l..

The main question that we want to address in this brief paper is if this
anomaly can be connected with a non-standard inflationary process.

As pointed in previous papers (see, for example, \cite{nid2012} and
references therein), a non zero chemical potential,
and, therefore, a larger value for $N_{eff}$ could arise in the curvaton 
scenario, proposed by \cite{Lyth:2001nq,Lyth:2002my}.

In this model, while the exponential expansion is driven by the inflaton field,
the primordial fluctuations are generated by a different field called "curvaton".
After the inflaton decay, the isocurvature perturbation produced initially by the curvaton
is converted in an adiabatic component. In this model some residual 
isocurvature perturbation is therefore expected in the cosmological fluids (cold dark matter,
baryons and neutrinos) (see, for example, \cite{Crotty:2003rz,Beltran:2004uv,Moodley:2004nz,Bean:2006qz,Trotta:2006ww,Komatsu:2010fb}).
In case of a non-vanishing neutrino chemical potential, neutrino density isocurvature perturbations are expected.

In few words, probing neutrino isocurvature density perturbation (NID hereafter), in the curvaton
scenario is complementary to constrain the lepton number in the neutrino sector.
It is therefore extremely timely to investigate the current CMB bounds on NID 
perturbation component, allowing at the same time a variation in the neutrino
effective number $N_{eff}$. 

Bounds on neutrino isocurvature perturbations have been presented in
the past in \cite{Gordon:2003hw} and \cite{Savelainen:2013iwa}. The Planck
collaboration has also provided new and stringent bounds on NID, but fixing 
$N_{eff}$ to the standard  value of $3.046$.

In this paper we present, for the first time, a combined analysis for $N_{eff}$
and NID from the Planck data, considering also the possibility of other
datasets as the recent Hubble constant measurements.

The paper is organized as follows. In Section \ref{nu} we review the NID perturbations 
which are generated in the curvaton scenario, in
Section \ref{analysis} we describe our analysis method, while in Section \ref{results}
we present the corresponding results. Our conclusions are reported in Sec.~\ref{Conclusions}.

\section{Neutrino isocurvature perturbations}\label{nu}

Let us remind the description of density perturbations in terms of the gauge invariant
variable $\zeta$ that describes the curvature perturbation on slices of uniform total density
\cite{Bardeen:1983qw,Mukhanov:1990me,Wands:2000dp}:

\begin{equation}
\zeta = -\psi - H \frac{\delta\rho}{\dot\rho} \, ,
\end{equation}

where the dot denotes derivatives with respect to the cosmological time $t$, $H$ is the Hubble parameter, $\psi$ is the (gauge-dependent) curvature perturbation, and  
$\rho$ the total energy density.

In the case of multiple fluids, it is possible to define the quantities $\zeta_i$ for each of the $i$-th energy component

\begin{equation}
\zeta_i = -\psi - H\frac{\delta\rho_i}{\dot\rho_i} \, .
\end{equation}

For an adiabatic mode the ratios $\delta\rho_i/\dot \rho_i$ are all the same, so that $\zeta_i = \zeta$ for all components. 
At the same time, an isocurvature fluctuation $\iso_i$ in the $i$-th energy 
component is given by the relative entropy fluctuation with respect to photons:
\begin{equation}
\iso_i\equiv 3(\zeta_i - \zeta_\gamma)\,.
\end{equation}

The relativistic neutrinos will follow an equilibrium distribution function as

\begin{equation}
f_i(E) = \left[\exp(E/T_\nu\mp\xi_i)\right]^{-1} \, \label{eq:FD}\,,
\end{equation}
where $T_\nu$ is their temperature, $\xi_i=\mu_i/T_\nu$ with $\mu_i$ as the chemical potential,
the index $i$ runs over the three neutrino families, $i=e,\,\mu,\,\tau$, and the minus (plus) sign is for neutrinos (antineutrinos). It is important to note that NID perturbations necessarily implies a non zero lepton asymmetry for the neutrino,  $n_L\equiv n_\nu - n_{\bar \nu}$, unless there is an exact cancellation of the asymmetries in the three flavours. 

Given the distribution function Eq. \reff{eq:FD}, the energy density $\rho_i\equiv \rho_{\nu_i}+\rho_{\bar\nu_i}$ in the 
high-temperature limit $T_\nu\gg m_\nu$ is given by \cite{Lesgourgues:1999wu}:
\begin{align}
&\rho_i = \frac{7\pi^2}{120} A_i \, T_\nu^4=\frac{7}{8}A_i \left(\frac{T_\nu}{T_\gamma}\right)^4 \rho_\gamma \; , 
\end{align}
where
\begin{align}
&A_i \equiv \left[1+\frac{30}{7}\left(\frac{\xi_i}{\pi}\right)^2+\frac{15}{7} \left(\frac{\xi_i}{\pi}\right)^4 \right] \; , \label{eq:Ai}
\end{align}

From our above definition, we have that  $\Neff = \sum_{i} A_i$. We can thus relate the isocurvature perturbation in the total neutrino density to the fluctuations $\delta\Neff^{(i)}$ (see \cite{nid2012}):

\begin{equation}
\iso_\nu=3(\zeta_\nu - \zeta_\gamma) \simeq \frac{\sum_i\delta\Neff^{(i)}}{4 \Neff}\,.\label{eq:isonu}
\end{equation}
In summary, a NID component is naturally connected to a non-standard value for $N_{eff}$.
In the next section we will therefore perform an analysis allowing both components to vary.

\section{Analysis method}\label{analysis}
Our analysis method is based on the Boltzmann CAMB
code~\cite{camb} and a Monte Carlo Markov Chain (MCMC) analysis based
on the MCMC package \texttt{cosmomc}~\cite{Lewis:2002ah}.

We sample the following set of parameters: 

\begin{equation}
\label{parameter}
  \{\omega_b,\omega_c, \Theta_s, \tau, n_s, \log[10^{10}A_{s}], N_{eff}, \alpha^{NID}\}~,
\end{equation}

$\omega_b\equiv\Omega_bh^{2}$ and $\omega_c\equiv\Omega_ch^{2}$  
being the physical baryon and cold dark matter energy densities,
$\Theta_{s}$ the ratio between the sound horizon and the angular
diameter distance at decoupling, $\tau$ is the reionization optical depth,
$n_s$ the scalar spectral index, $A_{s}$ the amplitude of the
primordial spectrum, $N_{eff}$ the effective neutrino number and
$\alpha^{NID}$ is the NID amplitude defined such that the total
CMB power spectrum is given by:

\begin{eqnarray}
 C_\ell&=& (1-\alpha^{NID})C_\ell^{ad} + \alpha^{NID} C_\ell^{nid}+\nonumber\\
&&+2 sign(\alpha^{NID}) \sqrt {\alpha^{NID}(1-\alpha^{NID})}C_\ell^{corr} \, ,
\label{eqn_cl}
\end{eqnarray}

where $C_\ell^{ad}$ is the adiabatic component, $C_\ell^{nid}$ is the
neutrino isocurvature density component and $C_\ell^{corr}$ is the correlated
spectrum. With this convention, when $\alpha^{NID} <0$ the spectra are totally anti-correlated.

These theoretical power spectra are then compared with the recent CMB
measurements made by the Planck experiment. For the Planck data, we add the 
high-$\ell$ and low-$\ell$ TT likelihoods and we also add the 
low-$\ell$ TE, EE, BB WMAP likelihood, see Ref.~\cite{Ade:2013lta} for
details. This corresponds exactly to the Planck+WP case presented in Ref.~\cite{Ade:2013lta}.
Moreover, we have marginalized over all foregrounds parameters, 
using the same procedure and priors presented in Ref.~\cite{Ade:2013lta}.
We also consider the HST constraint on the Hubble constant from
\cite{hst}.

\section{Results} \label{results}

\begin{table*}
\begin{center}
\begin{tabular}{|c|c|c|}
\hline\hline
Parameter & Planck+WP & Planck+WP+HST \\
\hline\hline
$\Omega_bh^2$ &$0.02215\pm0.00050$ &$0.02260\pm0.00033$ \\
$\Omega_{\rm c}h^2$ &$0.1222\pm0.0068$ &$0.1273\pm0.0056$ \\
$\theta$ &$1.0405\pm0.0010$ &$1.0408\pm0.0011$ \\
$\tau$ &$0.094\pm0.015$ &$0.099\pm0.015$ \\
$n_s$ &$0.966\pm0.021$ &$0.987\pm0.012$  \\
$log[10^{10} A_s]$ &$3.115\pm0.035$ &$3.122\pm0.037$ \\
$H_0 [\mathrm{km}/\mathrm{s}/\mathrm{Mpc}]$& $68.7\pm3.9$ &$72.5\pm2.2$\\
$N_{\rm eff}$ &$3.26\pm0.48$ &$3.70\pm0.30$ \\
$\alpha^{NID}$ &$-0.0031\pm0.0053$ &$0.0002\pm0.0031$ \\

\hline\hline
 \end{tabular}
 \caption{Constraints at $68 \%$ confidence level on $N_{eff}$, $\alpha^{NID}$ and
 the main $6$ cosmological parameters from the Planck+WP and Planck+WP+HST cases.}
 \label{table2}
 \end{center}
 \end{table*}

\begin{figure}[htb!]
%\centering
\includegraphics[width=\columnwidth]{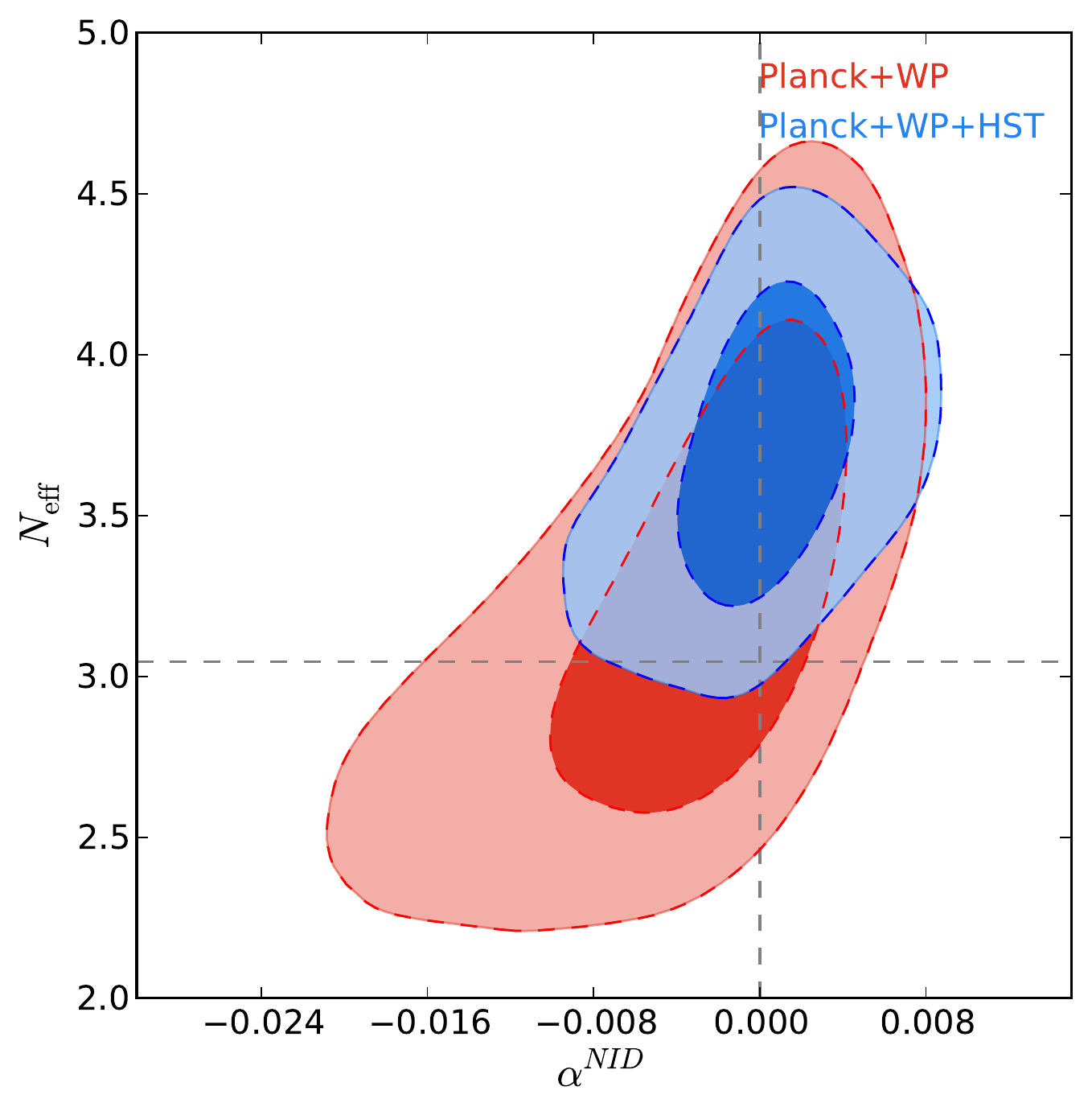} 
\caption{68\% and 95\% c.l.\ likelihood contours for Planck+WP and Planck+WP+HST
in the $N_{eff}$ vs. $\alpha^{NID}$ plane. Note the small correlation between 
the two parameters.} 
\label{gamma}
\end{figure}

The results of our analysis are reported in Table 1 and Figure 1, 
in the case of the Planck+WP and the Planck+WP+HST datasets.
As we can see, the Planck+WP data does not show any indication for
NID or for a larger value for $N_{eff}$. In practice, a cosmological
degeneracy exists along the $\alpha^{NID}$-$N_{eff}$ direction and
models with smaller values for $N_{eff}$ are more consistent with the
CMB observations when $\alpha^{NID}<0$. The current Planck+WP data alone 
does not show any supporting evidence for NID when variations in
$N_{eff}$ are considered. Moreover, the standard value of $N_{eff}=3.046$
is more consistent with Planck observations, due to the larger error
on this parameter when NID are considered.

The conclusion is slightly different when also the HST dataset is
included. As we can see, including HST reduces the error bars 
on the NID component while providing an indication for a non-standard
value for $N_{eff}$ at more than two standard deviations. Again, this is
consistent with the  anti-correlation between $N_{eff}$ and $\alpha^{NID}$,
mentioned above (see Figure 1).

\begin{table*}
\begin{center}
\begin{tabular}{|c|c|c|c|c|}
\hline\hline
Parameter & Planck+WP & Planck+WP+HST & Planck+WP & Planck+WP+HST \\
 & $\alpha^{NID}>0$ & $\alpha^{NID}>0$ & $\alpha^{NID}<0$ & $\alpha^{NID}<0$ \\
\hline
$\Omega_bh^2$ &$0.02260\pm0.00043$ &$0.02271\pm0.00031$ &$0.02198\pm0.00043$ &$0.02249\pm0.00031$ \\
$\Omega_{\rm c}h^2$ &$0.1287\pm0.0059$ &$0.1295\pm0.0050$ &$0.1196\pm0.0056$ &$0.1248\pm0.0049$\\
$\theta$ &$1.04149\pm0.00082$ &$1.04149\pm0.00082$ &$1.04012\pm0.00085$ &$1.04003\pm0.00080$ \\
$\tau$ &$0.095\pm0.014$ &$0.096\pm0.014$ &$0.093\pm0.014$ &$0.102\pm0.015$\\
$n_s$ &$0.987\pm0.017$ &$0.992\pm0.011$ &$0.958\pm0.018$ &$0.982\pm0.011$\\
$log[10^{10} A_s]$ &$3.100\pm0.033$ &$3.104\pm0.031$ &$3.119\pm0.034$ &$3.145\pm0.033$\\
$H_0 [\mathrm{km}/\mathrm{s}/\mathrm{Mpc}]$& $72.4\pm3.4$ &$73.3\pm2.0$ & $67.3\pm3.3$ &$71.8\pm2.0$ \\
$N_{\rm eff}$ &$3.71\pm0.42$ &$3.81\pm0.27$ &$3.08\pm0.40$ &$3.59\pm0.27$\\
$\alpha^{NID}$ &$<0.0023$ &$<0.0025$ &$>-0.0056$ &$>-0.0023$\\

\hline\hline
 \end{tabular}
 \caption{Constraints at $68 \%$ confidence level on $N_{eff}$, $\alpha^{NID}$ and
 the main $6$ cosmological parameters from the Planck+WP and Planck+WP+HST cases.
 The two cases $\alpha^{NID}>0$ and $\alpha^{NID}<0$ are considered separately.}
 \label{table}
 \end{center}
 \end{table*}

\begin{figure}[htb!]
%\centering
\includegraphics[width=\columnwidth]{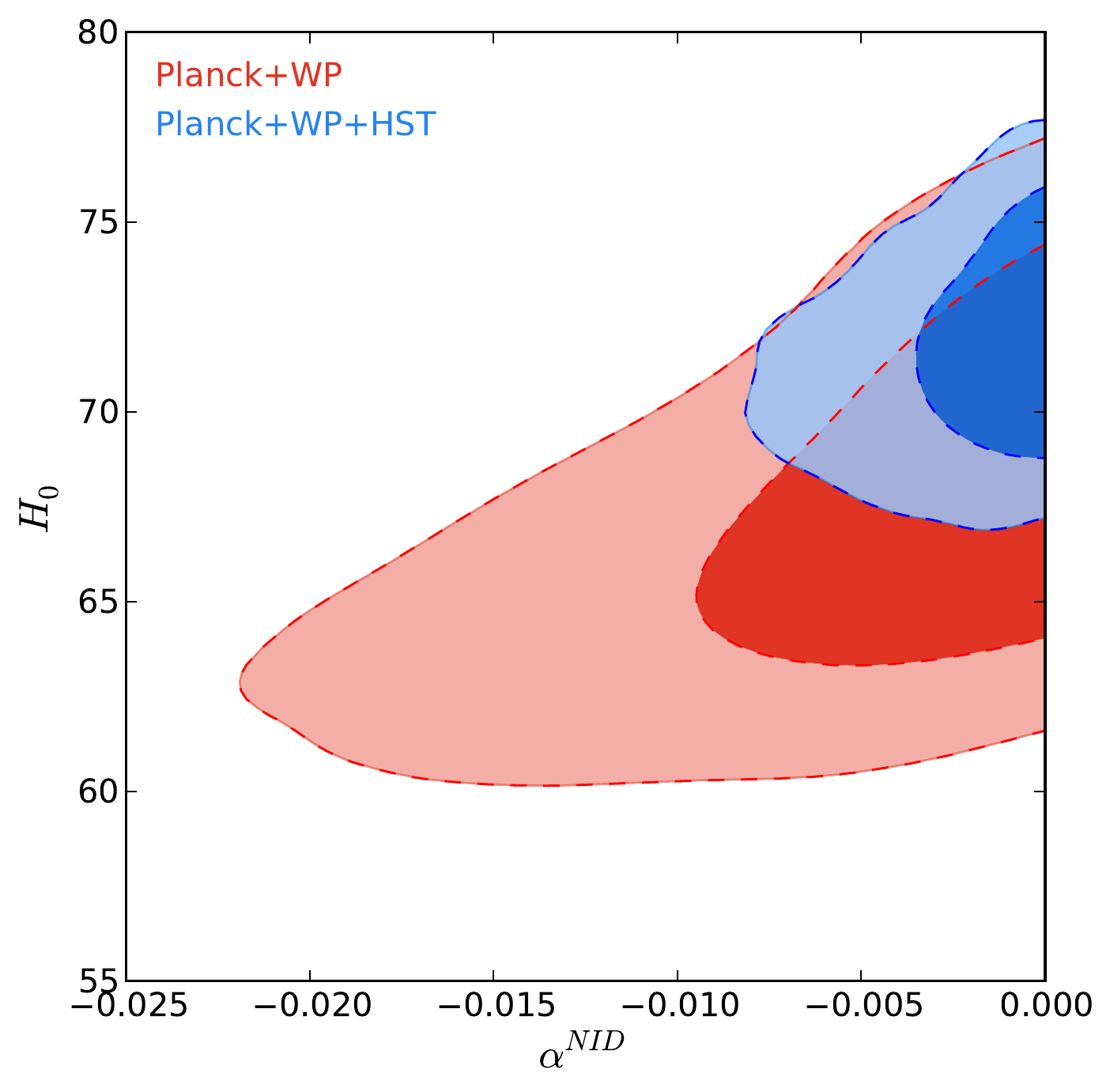} 
\includegraphics[width=\columnwidth]{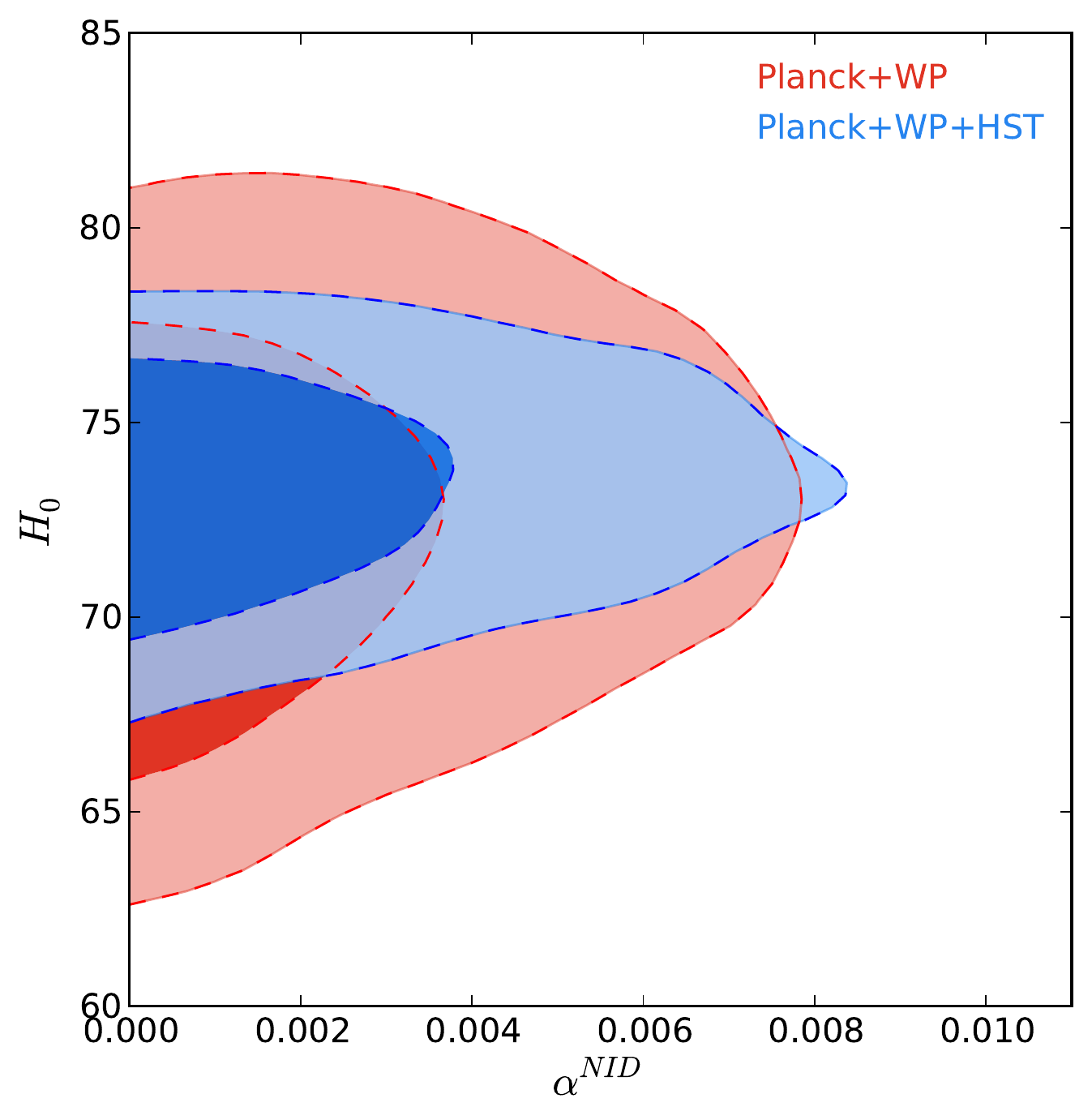} 
\caption{68\% and 95\% c.l.\ likelihood contours for Planck+WP and Planck+WP+HST.} 
\label{gamma}
\end{figure}

Since, as discussed in the previous section, a positive or a negative
value for $\alpha^{NID}$  discriminates between very different 
physical mechanisms for this NID component, it is interesting to repeat
the analysis but imposing each time the $\alpha^{NID}>0$ or 
$\alpha^{NID}<0$ prior.
The results for this analysis are reported in Table 2, for the
two datasets: Planck+WP and Planck+WP+HST.

As we can see, the interesting aspect is that when a $\alpha^{NID}<0$
prior is imposed, the Planck+WP case provide a value for the Hubble
constant that is in tension with current HST determinations, even
if the $N_{eff}$ parameter is allowed to vary.
It is clear from this that a NID component with $\alpha^{NID}<0$ 
can't resolve the  current tension on the values of $H_0$ between
the Planck data and the HST constraint.
On the other hand, the HST prior is clearly against a $\alpha^{NID}<0$
component, since including it drastically improves the constraint on this
parameter. 

In the case of $\alpha^{NID}>0$, on the contrary, the constraint on
the NID component are practically left unaffected by the inclusion of a
HST prior. This is evident from Figure 2, where we report the 2-D constraints on
the $H_0$ vs $\alpha^{NID}$ in the case of $\alpha^{NID}<0$ (Top Panel) and
$\alpha^{NID}>0$ (Bottom Panel) for the Planck+WP and Planck+WP+HST datasets.

\section{Conclusions}\label{Conclusions}

The recent Cosmic Microwave Background data from the Planck satellite experiment,
when combined with HST determinations of the Hubble constant, are compatible 
with a larger, non-standard, number of relativistic degrees of freedom at recombination,
 parametrized by the neutrino effective number $N_{eff}$. 
In the curvaton scenario, a larger value for $N_{eff}$ could arise from a non-zero
neutrino chemical potential connected to residual isocurvature perturbations after the
decay of the curvaton field, which component is parametrized by the
amplitude $\alpha^{NID}$. Here we present constraints on a joint analysis
of $N_{eff}$ and $\alpha^{NID}$. We found that the Planck+WP dataset does not
show any indication for a neutrino isocurvature component and that current
indications for a non standard $N_{eff}$ component are further relaxed.
When the HST prior on the Hubble constant is included, an anticorrelated,
$\alpha^{NID}<0$, neutrino isocurvature density component is severly constrained,
while the combined analysis suggests a value for $N_{eff}$ larger than the standard
expectations at more than two standard deviations.

%%%%%%%%%%%%%%%%%%%%%%%%%%%%%%%%%%%%%%%%%%%%%%%
\section*{Acknowledgments}
%%%%%%%%%%%%%%%%%%%%%%%%%%%%%%%%%%%%%%%%%%%%%%%
We are happy to thank Massimiliano Lattanzi, Gianpiero Mangano,
Pasquale Serpico, and Matteo Martinelli for useful discussion and help.

%%%%%%%%%%%%%%%%%%%%%%%%%%%%%%%%%%%%%%%%%%%%%%%

\end{document}